\documentclass[%
 reprint,
 amsmath,amssymb,
 aps,
prd,
]{revtex4-2}

\usepackage{graphicx}
\usepackage{dcolumn}
\usepackage{bm}
\usepackage{amsmath}
\usepackage{hyperref}

\numberwithin{equation}{section} 

\begin{document}

\preprint{APS/123-QED}

\title{Fluctuation-dissipation relation of test masses in classical stochastic gravitational wave background }
\author{Manjia Liang$^{1,2}$}
\email[Corresponding author:]{liangmanjia21@mails.ucas.ac.cn}

\author{Peng Xu$^{1,3,4,5}$}
\email[Corresponding author:]{xupeng@imech.ac.cn}
\author{Congfeng Qiao$^{2,3}$}
\author{Minghui Du$^{1}$}
\author{Qiong Deng$^{1}$}
\author{Bo Liang$^{1}$}
\author{Ziren Luo$^{1}$}

\affiliation{$^{1 }$Center for Gravitational Wave Experiment, National Microgravity Laboratory, Institute of Mechanics,
Chinese Academy of Sciences, Beijing 100190, China}
\affiliation{$^{2}$University of Chinese Academy of Sciences, Beijing 101400, China}

\affiliation{$^{3 }$Taiji Laboratory for Gravitational Wave Universe (Beijing/Hangzhou), UCAS, Beijing 100049, China}
\affiliation{$^{4 }$School of Fundamental Physics and Mathematical Sciences, Hangzhou Institute for Advanced Study, UCAS, Hangzhou 310024, China}

\affiliation{$^{5 }$Lanzhou Center of Theoretical Physics, Lanzhou University, Lanzhou 730000, China}

\date{\today}

\begin{abstract}
Research on pulsar timing arrays has provided preliminary evidence for the existence of a stochastic gravitational background, which, either being primordial or of astrophysical origin, will interact universally with matter distributions in our universe and affect their evolutions. This work, based on general relativity and stochastic dynamics theory, investigates the fluctuation-dissipation relation of isolated celestial bodies within a classical stochastic gravitational wave background. We employ the generalized Langevin model to analyze the fluctuating forces exerted on test masses by random spacetime and how energy dissipation occurs. Through the assumption of equilibrium, we derive the necessary conditions that should be satisfied by the stochastic gravitational wave background in the long wavelength limit, which, as we found, is a result of the back-reactions of the test mass system to the stochastic field. Additionally, as the establishment of the fluctuation-dissipation relation for such a system, certain thermodynamic quantities related to the statistical properties of the stochastic gravitational wave background could be defined and the characteristic of the diffusion process of test masses is obtained.

\end{abstract}

\maketitle

\section{Introduction}

The study of random motions can date back to the 19th century, beginning with the first detailed analysis of the motions of pollen particles in solution by the Scottish botanist Robert Brown in 1827. In the early 20th century, Sutherland, Einstein, and von Smoluchowski provided a theoretical explanation for these observations, pointing out that Brownian motions are caused by quasi-random microscopic interactions with the molecules that constitute the liquid. Einstein introduced the Einstein relation, laying the groundwork for the fluctuation-dissipation theorem that could be taken as the central result in the studies of fluctuations \cite{Einstein1905}. Building on Einstein's idea, Jean Baptiste Perrin had then precisely given the estimation of Avogadro's constant \cite{1967princeton}.

Einstein's contributions also paved the way for Langevin's approach to Brownian motion. In 1908, Langevin established a theoretical framework based on the theory of stochastic differential equations, incorporating both the inertia (damping) effect and the fluctuating force effect \cite{Langevin1908, Noelle2010, Uhlenbeck1930}. Such an approach clearly describes the disturbances experienced by Brownian particles due to random interactions with solvent molecules. 
The mathematical treatment of Brownian motion and the picture of almost everywhere continuous but nowhere differentiable trajectories were then established by N. Wiener in the 1920s \cite{Wiener1923}.
Such an idea that evolutions of thermodynamic observables are driven by fluctuations, as described by Langevin equations, led to the modern linear response theory of non-equilibrium processes \cite{2017Nonequilibrium}.

In 1985, Ford demonstrated that the stochastic dynamics of charged particles in a blackbody radiation (BBR) field could be described using the model of independent harmonic oscillators, reformulating the equation of motions in the form of a generalized Langevin equation \cite{Ford1988, Bao2005, Ford1985, Connell2010}. 
The generalized Langevin equation, with modifications of retarded frictions,  was established to study anomalous diffusion behavior driven by nonwhite noise fluctuations. The anomalous diffusion behavior is connected with the breakdown of the central limit theorem that is caused by either broad distributions or long-range correlations \cite{1990Anomalous, Hughes1995}.
When discussing the nonlinear growth of the mean square displacement (MSD) over time, the anomalous diffusion exponent $\alpha$ to describe different types of anomalous diffusion behavior, including superdiffusions ($0< \alpha < 1$) and subdiffusions ($\alpha > 1$). 
In the 1990s, the fractional dynamics approach to anomalous diffusion was introduced by Ralf Metzler and Joseph Klafter \cite{2000The}.

In 1915, after the establishment of the general theory of relativity (GR), Einstein predicted the existence of Gravitational Waves (GW). A century later, in 2015, the first and groundbreaking direct observation of a GW signal from the merger of a black hole binary system was obtained by the LIGO-Virgo collaboration \cite{Abbott2016}. This discovery marked a major milestone for the theoretical study and experimental investigations of GR, and together with the follow-on observations, ushered in a new era of astrophysics and cosmology. 

In addition to isolated sources like coalescing binaries and rotating neutron stars, the stochastic gravitational wave background (SGWB), whether of primordial or astrophysical origin, is also of great significance.
Primordial SGWB may come from phase transitions or turbulences in matter distributions in the early stage of the universe \cite{Allen1988}. 
Astrophysical original SGWB is generally formed by superpositions of GWs from numerous isolated sources, such as binary systems. 
In 2023, research teams from the National Astronomical Observatories of the Chinese Academy of Sciences and the North American Nanohertz Observatory for Gravitational Waves established the key evidence of the existence of a GW background in the nanohertz band, respectively \cite{2023Searching, agazie2023nanograv}.As a cosmological background, it is believed that the SGWB should not be taken as some isolated background field \cite{Vishnu2024}, but may leave fingerprints in the matter distributions during the lengthy evolution of our universe.

This matter can be understood well through the imagery of Brownian motion.
If we liken celestial bodies to pollen, then the SGWB would act as the solvent, providing the foundation for the random motion of these bodies. 
The statistical properties of this random motion are determined by the statistical properties of the SGWB.
Besides, the objects that undergo random motion driven by the background will, in turn, impact the SGWB. 
These two processes are greatly related to the concepts of fluctuations and dissipation in non-equilibrium statistics.
Inspired by this idea, the purpose of this article is to study this fluctuation-dissipation process, which results from the randomness of SGWB.
To achieve this, we need to establish two very important physics processes at first: how the test mass dissipates energy to the SGWB and how the SGWB drives the test mass to undergo random motion. 
Then we make an equilibrium assumption and derive the necessary conditions for the equilibrium state of the SGWB. The results we have obtained are consistent with the second law of thermodynamics.
Based on the fluctuation-dissipation relation, we derive some well-defined thermodynamic statistical quantities for the SGWB.

The macroscopic statistical properties of SGWB could be inspiring for understanding the microscopic picture of gravity.
This is akin to how the thermodynamics of electromagnetic fields plays a crucial role in the study of quantization of classical electromagnetic theory, with achievements like Planck's blackbody radiation formula and the equation of state for radiation fields having widespread applications in physics \cite{Tolman1934}. Although, to this day, there have been many outstanding achievements in studying the thermodynamics of SGWB  \cite{freidel2015gravitational, christensen2018stochastic, agullo2013extension}, the fluctuation-dissipation relation resulting from SGWB has not yet been established.

Our research begins by considering GWs as a type of classical wave, since, even today, there is no universally accepted theory of Quantum Gravity.
The specific structure of the paper is as follows.
In Sec. II, We will briefly introduce the Langevin equation and its microscopic interpretation in classical diffusion models. 
In Sec. III, We derive the friction coefficient and the fluctuating force based on the random properties of the SGWB in Minkowski spacetime. We will present the generalized Langevin equation that drives the evolution of random particles and the conditions imposed on the background spectral density function if we require that all Brownian particles eventually reach equilibrium with the background. 
In Sec. IV, we summarize our results.

\section{Classic Diffusion Model}

\subsection{The Langevin model}

In the absence of external potentials, a Brownian particle is said to be ``free'', and for the one-dimensional (1D) case the Langevin equation for the random motion reads
\begin{equation}
M\frac{dv}{dt}=-M\gamma v+F\left ( t\right ),
\end{equation}
where $M$ is the mass of the Brownian particle and $v$ is the velocity. 
The Langevin equation is historically the first example of a stochastic differential equation, that is, a differential equation involving a random term $F\left ( t\right )$ with specified statistical property.
The term $-M\gamma v$ is the inertia or damping term, with the relaxation time $\tau _{\gamma }=\frac{1}{\gamma}$ for the diffusion process.
The random term $F\left ( t\right )$, which accounts for the fluctuating force acting on the Brownian particle, is assumed to have an average value of zero for simplicity, 
\begin{equation}
\left \langle F\left ( t\right )\right \rangle=0.
\end{equation}
The angle brackets here represent statistical average operations. Without external potentials, this condition is necessary to provide a vanishing average velocity at equilibrium. For stationary fluctuating forces  $F\left ( t\right )$, the statistical properties can be described by the autocorrelation function $g\left ( \tau \right )$ and fluctuation parameter $\mathcal{D}$,
\begin{gather}
g\left ( \tau \right )=\left \langle F\left ( t\right )F\left ( t+ \tau \right )\right \rangle,
\\\int_{-\infty }^{\infty }g\left ( \tau \right )d\tau =2\mathcal{D}M^{2}.
\end{gather}
The Langevin equation is generally applied to systems with $g\left ( \tau \right )$ being a $\delta $ function.

\subsection{Friction coefficient and fluctuation parameters of classical diffusion models}

In this section, we will briefly introduce two classic Brownian motion models: (1) the random motion of macromolecules in a solvent composed of point particles  \cite{Bogolyubov1945, Jörn2006}, and (2) the random motion of a charged particle in a BBR field.

\subsubsection{Motion of macromolecules in solvents}

For simplicity, we will restrict our discussion to the one-dimensional (1D) scenario. Consider the following situation in an inertial laboratory frame: a large 1D box
$\mathcal{V}=\left [-L/2,L/2\right ]$ contains an ideal gas at temperature $T$, consisting of $N$ small and pointlike particles, each with identical mass $m$. 
The system of the gas particles, referred to hereafter as the heat bath, surrounds a Brownian particle of mass $M$. We assume that the interactions between the thermal background and the Brownian particle are of elastic collisions.
 
According to the conservations of momentum and energy, the change in the Brownian particle’s momentum due to each elastic collision is given by:
\begin{equation}
\Delta P=\frac{-2m}{M+m}P+\frac{2M}{M+m}p,
\end{equation}
where $P$ is the momentum of the Brownian particle and $p$ of the gas particles.

Let us define ${I}_{r}(t,\tau )$  as the indicator function representing a collision between the $r-th$ heat bath particle and the Brownian particle during the interval $\left [ t,t+\tau \right ]$. This function can be explicitly written as:
\begin{equation}
\begin{aligned}{I}_{r}(t,\tau )
 &=\Theta  (X-{x}_{r})\Theta ({x}_{r}{}^{\prime}-{X}^{\prime})\Theta ({v}_{r}-V) \\ 
 &\; +\Theta  ({x}_{r}-X)\Theta ({X}^{\prime}-{x}_{r}{}^{\prime})\Theta (V-{v}_{r}),
\end{aligned}
\end{equation}
where
\begin{equation}
\begin{matrix}{X}^{\prime}=X+V\tau,
 & {{x}_{r}}^{\prime}={x}_{r}+{v}_{r}\tau ,
\end{matrix}
\end{equation}
and the Heaviside function is defined by
\begin{equation}
\Theta (x)=\left\{\begin{matrix}0,
 & x<0\\ 1/2,
 & x=0\\ 1,
 & x>0.
\end{matrix}\right.
\end{equation}
The total momentum change $\delta P$ of the Brownian particle within the time interval $\tau $ can be written as
\begin{equation}
\begin{aligned}\frac{\delta P(t)}{\tau }=
 & -\left [ \frac{1}{\tau }\displaystyle\sum_{r=1}^{N}\frac{2m}{m+M}{I}_{r}(t,\tau )\right ]P\\ 
 & \quad \quad \quad \quad \quad +\frac{1}{\tau }\displaystyle\sum_{r=1}^{N}\frac{2M}{M+m}{p}_{r}{I}_{r}(t,\tau ).
\end{aligned}
\end{equation}
The first term on the right-hand side can be identified as the inertia term, while the second term is the fluctuating force $F(t)$.

So, we can write the friction coefficient and the fluctuating force as follows:
\begin{gather}
\gamma =\frac{1}{\tau }\displaystyle\sum_{r=1}^{N}\frac{2m}{m+M} {I}_{r}(t,\tau ),
\\F(t)=\frac{1}{\tau }\displaystyle\sum_{r=1}^{N}\frac{2M}{M+m}{p}_{r}{I}_{r}(t,\tau ).
\end{gather}
Since the heat bath particles are assumed to be independently and identically distributed, we can rewrite $\gamma$ as
\begin{equation}
\gamma =\frac{N}{\tau }\frac{2m}{m+M}\left \langle {I}_{r}(t,\tau )\right \rangle.
\end{equation}

\subsubsection{Motion of a charged particle in BBR field}

The collision model is relatively simple compared to the much more complex dynamics of charged particles in a BBR field. The answer to this question was first provided by Ford \cite{Ford1988}, who proposed the independent harmonic oscillator (IO) model. He described the motion of Brownian particles using the Hamiltonian equation of this model. In this framework, the charged particle is coupled with a bath of  $N$ independent harmonic oscillators, each with mass ${m}_{j}$ and intrinsic frequency ${\omega }_{j}$, described by the coordinates ${x}_{j}$ and the conjugate momenta ${p}_{j}$.
The Hamiltonian of the entire system can be expressed as
\begin{equation}
H=\frac{{p}^{2}}{2M}+\phi \left ( x\right )+\frac{1}{2}\displaystyle\sum_{j=1}^{N}\left [ \frac{{p}_{j}^{2}}{{m}_{j}}+{m}_{j}{\omega }_{j}^{2}({x}_{j}-x){}^{2}\right ],
\end{equation}
where $\phi \left ( x\right )$ is the external potential. In the case of a free particle, $\phi \left ( x\right )$ is set to zero

Based on such a Hamiltonian, we can obtain Hamilton's equations of Brownian particles,
\begin{equation}
M\frac{dv}{dt}=-M\int_{-\infty }^{t}\gamma \left ( t-{t}^{\prime}\right )v\left ( {t}^{\prime}\right )d{t}^{\prime}+F\left ( t\right ),
\end{equation}
where $\gamma \left ( t\right )$ and $F\left ( t\right )$ can be expressed as
\begin{equation}
\gamma \left ( t\right )=\frac{1}{M}\displaystyle\sum_{j}^{}{m}_{j}{\omega }_{j}^{2}cos\left ( {\omega }_{j}t\right )\Theta \left ( t\right ),
\end{equation}
\begin{equation}
\label{Ft}
F\left ( t\right )=\displaystyle\sum_{j}^{}{m}_{j}{\omega }_{j}^{2}{x}_{j}^{h}\left ( t\right ).
\end{equation}
In the equation, ${x}_{j}^{h}\left ( t\right )$ denotes the general solution of the homogeneous equation for the heat-hath oscillators (corresponding to no interaction). Substitute the mass ${m}_{j}$ of the harmonic oscillator:
\begin{equation}
{m}_{j}=\frac{4\pi {e}^{2}{f}_{j}^{2}}{{\omega }_{j}^{2} V},
\end{equation}
where ${f}_{j}$, $e$, and $V$ are the electron form factor, the charge of the Brownian particles, and the volume of the blackbody cavity, respectively. We can get the real part of the Fourier–Laplace transform of the memory kernel $\gamma \left ( t\right )$:
\begin{equation}
\mathfrak{R}e\left [ \gamma \left ( \omega \right )\right ]=\frac{1}{M}\frac{2{e}^{2}{\omega }^{2}}{3{c}^{3}}{f}_{j}^{2}.
\end{equation}
Additionally, Eq. (\ref{Ft}) can be further simplified to $F=eE$ based on quantum electrodynamics \cite{Connell2010}, where $E$ is the electric field intensity of the BBR field.

When the Brownian particle reaches equilibrium with the BBR field, the average power imparted to the Brownian particle by the BBR field can be expressed as $\left \langle vF(t)\right \rangle$. On the contrary, the average power given to the BBR field by Brownian particles can be expressed as \cite{Harko2016}
\begin{equation}
\left \langle P_{d}\right \rangle=\frac{2{e}^{2}}{3{c}^{3}}\left \langle {a}^{2}\right \rangle,
\end{equation}
where $a$ is the acceleration of the Brownian particle and this equation is known as the Larmor formula. Clearly, at equilibrium, $\left \langle vF(t)\right \rangle$ should be equal to $\left \langle P_{d}\right \rangle$.

\section{Brownian motion in SGWB}

Our goal in this section is to derive the Langevin equation for Brownian particles in SGWB. The main approach involves studying the momentum change between randomly moving particles and the SGWB. To achieve this, we decompose the spacetime metric into the flat background ${\eta }_{\mu \nu }$ and the superposition of GW perturbations ${h}_{\mu \nu  }\left ( t,\mathbf{x}\right )$,
\begin{equation}
{g}_{\mu \nu }={\eta }_{\mu \nu }+{h }_{\mu \nu }+\mathcal{O}\left ( {h}^{2}\right ).
\end{equation}
On this basis, the calculation of the properties of the specific stochastic process will be conducted within the reference frame that is at rest relative to the SGWB, in which the observed intensity of the SGWB is uniform in all directions. On a relatively small time scale compared to the age of the universe, we can neglect the effects of cosmic expansion and treat it as an inertial frame. Additionally, we will set the speed of light $c$ to 1.

\subsection{Fluctuating force on test mass in SGWB}
The motion of a test mass $M$ in a gravitational field is described by the geodesic equation,
\begin{equation}
\frac{{d}^{2}{x}^{\mu }}{d{s }^{2}}+ {\varGamma }_{\rho \sigma  }^{\mu }\frac{d{x}^{\rho }}{{d s }}\frac{d{x}^{\sigma  } }{{d s }}=0,
\end{equation}
where ${x}^{\mu }=\{t,x^i\},\ (i=1,2,3)$ are the coordinates of the test mass in the cosmological rest frame of the SGWB, and $s$ is its proper time.
The connection component ${\varGamma }_{\rho \sigma  }^{\mu }$ contains statistical information about the random spacetime fluctuations. 

It is natural to assume that the test mass satisfies the slow motion limit, and the 4-velocity of the test mass can be written as $u^{\mu}=\{1+\mathcal{O}({v}^{2}), v^i+\mathcal{O}({v}^{2})\}$, with $v^i=dx^i/dt$. 
Then the above geodesic equation along certain directions (take the $x^3$-direction as an example) can be written as
\begin{equation}
\label{cedi}
\left ( \frac{{d}^{2}{x}^{3 }}{d{t }^{2}}+ {\varGamma }_{\rho \sigma  }^{3}{u}^{\rho }{u}^{\sigma  }\right )(1+\mathcal{O}(v)+\mathcal{O}(h))=0.
\end{equation}
We can further simplify this formula using this relationship $\mathcal{O}(h)\mathcal{O}(v) \ll \mathcal{O}(h)$.
Then, Eq. (\ref{cedi}) becomes
\begin{equation}
\label{lianluo}
\frac{d{v}^{3}}{dt}=-{\varGamma }_{00}^{3}.
\end{equation}
Then, within the slow-motion limit, there left only one term ${\varGamma }_{00}^{3}$ from the spacetime disturbance in the geodesic equation, and $M{\varGamma }_{00}^{3}$ will appear as the fluctuating force in the generalized Langevin equation along $x^3$-direction.

In fact, the test mass experiences fluctuating forces in the $x^1$ and $x^2$-directions similarly. In the following, for clarity, we will at first work out the fluctuations and dissipation from the SGWB in the 1D case along the $x^3$ direction, and then extend the analysis to the full 3D case.

\subsection{Spectral properties of uniform and isotropic SGWB}

In the weak field approximation, we can write the connection component ${\varGamma }_{00}^{3}$ in Eq. (\ref{lianluo}) as
\begin{equation}
\label{gama}
{\varGamma }_{00}^{3}(t,\mathbf{x})=\frac{1}{2}({\eta }^{\lambda 3}-{h}^{\lambda 3} )(2{h}_{\lambda 0,0}-{h}_{00,\lambda }).
\end{equation}
In the equation, the metric components are random variables on the space-time coordinates.

For a random variable $h(t)$ at fixed $\mathbf{x}$, its Fourier transform $h(\omega )$ is also a random variable,
\begin{equation}
h(\omega )=\int_{-\infty }^{\infty }h(t){e}^{iwt}dt.
\end{equation}
This equation and also the  inverse Fourier transform hold true in the mean square sense,
\begin{equation}
h(t)=\frac{1}{2\pi }\int_{-\infty }^{\infty }h(\omega ){e}^{-iwt}d\omega ,
\end{equation}
where $h(t)$ and $h(\omega )$ are random variables in the time domain and frequency domain, respectively. If $h(t)$ under consideration being stationary, the average of $h(t)$ is a constant $h$, and the average of $h(\omega )$ can be given by 
\begin{equation}
\left \langle h(\omega )\right \rangle=\int_{-\infty }^{\infty }\left \langle h(t)\right \rangle{e}^{iwt}dt=h\delta (\omega ).
\end{equation}
For the case of SGWB, we can set $h=0$.
The auto-correlation function of $h(t)$ is defined as
\begin{equation}
\kappa (\tau  )=\left \langle {h}^{\dagger}(t+\tau  )h(t)\right \rangle.
\end{equation}
To make the auto-correlation function only dependent on the time interval $\tau $, $h(\omega )$ must satisfy the following relationships:
\begin{equation}
\left \langle {h}^{\dagger}({\omega }^{\prime})h(\omega )\right \rangle=2\pi \delta (\omega -{\omega }^{\prime})S(\omega ).
\end{equation}
$S(\omega )$ is the spectral density function of $h(t)$, that satisfies the following relationship:
\begin{equation}
\kappa (\tau  )=\frac{1}{2\pi }\int_{-\infty }^{\infty }S(\omega ){e}^{iw\tau }d\omega.
\end{equation}
This is known as the Wiener-Khintchine theorem.

Come back the connection defined in Eq. (\ref{gama}), which consists of the derivatives of the components of the metric, is obviously a random variable too. For random variables like $h(t)$, we can only define the mean square derivatives in the case that its autocorrelation function $\kappa (\tau )$ is a generalized second-order differentiable function.

If the mean square derivative of $h(t)$ exists, we have
\begin{equation}
{h}_{,0}(t)=\frac{1}{2\pi }\int_{-\infty }^{\infty }-i\omega h(\omega ){e}^{-iwt}d\omega .
\end{equation}
If $h(t)$ is stationary $\left \langle {h}_{,0}(t)\right \rangle=0$ since $\left \langle {h}_{,0}(t)\right \rangle={\left \langle h\left ( t\right )\right \rangle}_{,0}$. Furthermore, We can give an expression for the auto-correlation function of ${h}_{,0}(t)$,
\begin{equation}
\label{kapa}
{\kappa }_{d}(\tau )=\left \langle {h}_{,0}^{\dagger}(t+\tau{)}{h}_{,0}(t)\right \rangle
=\frac{1}{2\pi }\int_{-\infty }^{\infty }{\omega }^{2}S(\omega ){e}^{iw\tau }d\omega.
\end{equation}
Based on the Wiener-Khintchine theorem, the spectral density function of ${h}_{,0}(t)$ can be written as ${\omega }^{2}S(\omega )$.

For a function that depends on both time and space, we can decompose it as follows: \cite{Allen1999, Thrane2013}
\begin{equation}
{h}_{\mu \nu }\left ( t,\mathbf{x}\right )=\frac{1}{ 2\pi }\int_{-\infty }^{\infty }d\omega \int_{-\infty }^{\infty }d{}^{3}\mathbf{k} {h}_{\mu \nu }(\omega ,\mathbf{k}){e}^{-i\omega t+i\mathbf{k}\cdot \mathbf{x}},
\end{equation}
where $\mathbf{k}$ is the three-dimensional wave vector and must satisfy the condition ${\omega }^{2}={k}^{2}$. 
For a stationary and uniform SGWB, we have,
\begin{equation}
\left \langle {h}_{\mu \nu }\left ( t,\mathbf{x}\right )\right \rangle{,}_{\alpha }=0
\end{equation}
This leads to the conclusion that the means of the connection components should vanish, $\left \langle {\varGamma }_{\mu \nu }^{\rho }\left ( t\right )\right \rangle=0$. 
The spectral density function of ${h}_{\mu \nu }\left ( t,\mathbf{x}\right )$ can be written as
\begin{equation}
\left \langle {{h}^{\dagger}}_{\mu \nu }(\omega {}^{\prime},\mathbf{k}{}^{\prime}){h}_{\mu \nu }(\omega ,\mathbf{k})\right \rangle=(2\pi ){\delta }^{3} ({\mathbf{k}}^{\prime}-\mathbf{k})\delta ({\omega }^{\prime}-\omega )S_{\mu \nu }(\omega ,\mathbf{k}).
\end{equation}
By considering the isotropic hypothesis, we have
\begin{equation}
S_{\mu \nu }(\omega )=4\pi S_{\mu \nu }(\omega ,\mathbf{k}).
\end{equation}
Then we can write the autocorrelation function of the mean square derivative of ${h}_{\mu \nu }(t,\mathbf{x})$ 
\begin{equation}
\left \langle ({h}_{\mu \nu }^{,i}){}^{\dagger }(t+\tau ,\mathbf{x}){h}_{\mu \nu }^{,i}(t,\mathbf{x})\right \rangle=\frac{1}{2\pi }\int_{-\infty }^{\infty }\frac{{\omega }^{2}}{3}S_{\mu \nu }(\omega ){e}^{i\omega \tau }d\omega ,
\end{equation}
\begin{equation}
\left \langle ({h}_{\mu \nu }^{,0}){}^{\dagger }(t+\tau ,\mathbf{x}){h}_{\mu \nu }^{,0}(t,\mathbf{x})\right \rangle=\frac{1}{2\pi }\int_{-\infty }^{\infty }{\omega }^{2}S_{\mu \nu }(\omega ){e}^{i\omega \tau }d\omega .
\end{equation}
In the first equation, we have used the condition $\left \langle{k}_{i}^{2} \right \rangle={\omega }^{2}/{3}$.

At the same time, we can use the harmonic coordinate condition,
\begin{equation}
\overline{h}{}_{\mu \alpha }^{, \alpha }(t,\mathbf{x})=0,
\end{equation}
where,
\begin{equation}
\overline{h}{}_{\mu \nu }={h}_{\mu \nu }-\frac{1}{2}{\eta }_{\mu \nu }{h_{\lambda }}^{\lambda }.
\end{equation}
With this condition, We can reduce the degrees of freedom of the metric to 6, which are ${h}_{ij}$. By using the assumption of homogeneity of space, the six random variables ${h}_{ij}$ should satisfy the same statistical properties, which means that they are the same in the sense of random variables and should have the same mean and high-order moments. So we can set their spectral density functions as follows,
\begin{equation}
{S}_{ij}\left ( \omega \right )=S\left ( \omega \right ).
\end{equation}
This means that the spectral density functions of an isotropic and homogeneous SGWB are equivalent to having only one “degree of freedom.” 

Based on Eq. (\ref{gama}), we have
\begin{equation}
\label{ga}
{\varGamma }_{00}^{3}\left ( \omega ,\mathbf{k}\right )=-i\left (\omega {h}_{30}+\frac{1}{2}{k}_{3}{h}_{00} \right ).
\end{equation}
By using the third component of harmonic coordinate condition, which can be written as
\begin{equation}
\omega {\overline{h}}_{30}\left ( \omega ,\mathbf{k}\right )+{k}_{1} {\overline{h}}_{31}\left ( \omega ,\mathbf{k}\right )+{k}_{2} {\overline{h}}_{32}\left ( \omega ,\mathbf{k}\right )+{k}_{3} {\overline{h}}_{33}\left ( \omega ,\mathbf{k}\right )=0,
\end{equation}
we can rewrite Eq. (\ref{ga}) as 
\begin{equation}
\label{sf}
{\varGamma }_{00}^{3}\left ( \omega ,\mathbf{k}\right )=i\left ( {k}_{1}{h}_{13}+{k}_{2}{h}_{23}\right )-\frac{i}{2}{k}_{3}\left ({h}_{11}+{h}_{22}-{h}_{33} \right ).
\end{equation}
Then, we can get the spectral density function of the fluctuation force $F^3 $ along the $x^3$-direction from the uniform and isotropic SWGB,
\begin{equation}
{S}_{F^3}\left ( \omega \right )=\frac{11}{12}{M}^{2}S\left ( \omega \right ){\omega }^{2}.
\end{equation}
The form of the spectral density function of $F^3 $ depends on our chosen coordinate, and there is no covariance. This is because the definitions of the connection components depend on the coordinate system's choice. However, the contraction of the metric remains invariant under coordinate transformations. By calculating ${h}_{\mu \nu }{h}^{\mu \nu }$, we can obtain the spectral density function of the gauge-invariant statistical quantities of SGWB. So, the spectral density function of ${h}_{\mu \nu }$ can be written in a form with covariance.

Based on the assumption of homogeneity of space, the fluctuation forces in the other two directions have the same spectral density function as $x^3$-direction. For a complete 3D random motion, its fluctuation forces are described by $\vec{F}$. Its spectral density function is the sum of the spectral density function of the components in three directions
\begin{equation}
{S}_{\left | \vec{F}\right |}\left ( \omega \right )=\frac{11}{4}{M}^{2}S\left ( \omega \right ){\omega }^{2}.
\end{equation}

\subsection{Gravitational dissipation}

It is known that, for an isolated and dynamical matter system, ational radiation starts from the second time derivatives of the mass quadrupole moment. 
This is because the dipole moment of the isolated matter system gives rise to the center of mass, which will move at a uniform speed and have vanishing second-time derivatives.
According to GR, gravitational radiation dominated by quadrupole radiation will introduce radiation reaction force to the motion of celestial bodies, thereby leading to changes in their orbital motion.
However, the situation will be a little different for the problem considered in this work. To calculate the power of energy dissipation due to radiation reaction force on the test mass in the SGWB, denoted as $\left \langle {P}_{GD}\right \rangle$, where GD represents gravitational dissipation, one has to understand that $\left \langle {P}_{GD}\right \rangle$ represents the averaged energy that the test mass gives to the background in unit time, where the background contains all the substances in space except for the test mass under considerations.

When computing the self-gravitational field caused by the motion of the test mass itself, we can proceed under the Lorenz-gauge and the weak field approximation \cite{Barack2019}.
As in the previous subsection, it is assumed that the test mass $M$ is initially located near the origin and moves slowly along the $x^3$-axis with velocity $v$. For a point particle, the energy-momentum tensor reads
\begin{equation}
{T }_{\mu \nu}\left ( \mathbf{x},t\right )=M{u}_{\mu }{u}_{\nu }\delta{}^{3} \left ( \mathbf{x}-\mathbf{z}(t)\right ),
\end{equation}
where $\mathbf{z}(t)$ is the trajectory of the test mass.
In this case, the perturbations ${h}_{\mu \upsilon }$ are described by the linearized Einstein field equations \cite{Weinberg1972,Maggiore2007},
\begin{equation}
{\Box }^{2}{h}_{\mu \upsilon }=-16\pi G{\mathcal{S}}_{\mu \nu },
\end{equation}
where 
\begin{equation}
{\mathcal{S}}_{\mu \nu }={T}_{\mu \nu }-\frac{1}{2}{\eta }_{\mu \nu }T_{\lambda }\,^{\lambda }.
\end{equation}
One solution to this equation is the retarded solution,
\begin{equation}
{h}_{\mu \nu }(\mathbf{x},t)=4G\int {d}^{3}\mathbf{x}^{\prime}\frac{{\mathcal{S}}_{\mu \nu }(\mathbf{x}^{\prime},t-\left | \mathbf{x}-\mathbf{x}^{\prime}\right |)}{\left | \mathbf{x}-\mathbf{x}^{\prime}\right |},
\end{equation}
Where $\mathbf{x}^{\prime}$ represents the position of the source and only takes values within the source distribution region.
At the location very far from the source, $\mathbf{x}\gg \mathbf{x}^{\prime}$, we can replace $\left | \mathbf{x}-\mathbf{x}^{\prime}\right |$ with $r$ $(r=\left | \mathbf{x}\right |)$.

Then we can work out the solutions of ${h}_{\mu \nu }$, and non-zero terms are ${h}_{00},{h}_{11},{h}_{22},{h}_{33}$ and ${h}_{03}$. 
${h}_{03}$ can be written as
\begin{equation}
\begin{aligned}{h}_{03}(\mathbf{x},t)
 &=4G\int {d}^{3}{x}^{\prime}\frac{{T}_{03}({\mathbf{x}}^{\prime},t-r)}{r} \\ 
&=\frac{4G}{r}Mv^*
\end{aligned}
\end{equation}
where $v^*$ is the retarded speed. If the test mass is isolated, it must satisfy ${{h}}_{03,0}(\mathbf{x},t)=0$. However, in the case we studied, we can define the acceleration $a$ of the test mass driven by the disturbances from the SGWB with respect to the rest frame. 
As for other non-zero terms, we can also write out their specific form:
\begin{equation}
\begin{aligned}{h}_{ 33}(\vec{x},t)
 & =4G\int {d}^{3}{x}^{\prime}\frac{{\mathcal{S}}_{33}({\vec{x}}^{\prime},t-\left | \vec{r}\right |)}{\left | \vec{r}\right |}\\ 
 & =\frac{2G}{r}M\left ( v{}^{*}\right ){}^{2}+\frac{2G}{r}M,
\end{aligned}
\end{equation}
\begin{equation}
{h}_{ 00}(\vec{x},t)=\frac{2G}{r}M\left ( v{}^{*}\right ){}^{2}+\frac{2G}{r}M,
\end{equation}
\begin{equation}
{h}_{ 11}(\vec{x},t)=\frac{2G}{r}M-\frac{2G}{r}M\left ( v{}^{*}\right ){}^{2},
\end{equation}
\begin{equation}
{h}_{ 22}(\vec{x},t)=\frac{2G}{r}M-\frac{2G}{r}M\left ( v{}^{*}\right ){}^{2}.
\end{equation}
Then, one has $h_{\mu\mu,0}\sim\left ( GM v{}^{*}a{}^{*}/r\right ),\ (\mu=0,1,2,3)$. 
When the velocity $v$ is a small quantity, ${{h}}_{\mu\mu,0}\sim \mathcal{O}(h_{03,0}v)$ is much smaller than ${{h}}_{03,0}$.

After we get the specific form of ${h}_{\mu \nu }$, we will calculate the corresponding dissipated energy through the average energy-momentum tensor of GWs. The power with which matter dissipates energy per unit of solid angle in a certain direction can be expressed as follows \cite{Weinberg1972}:
\begin{equation}
\label{dP}
\begin{aligned}\frac{dP}{d\Omega }(\mathbf{x },\omega)=
 &  \frac{{r}^{2}{\omega }^{2}}{32\pi G}[ {h}^{\mu \nu \dagger }(\mathbf{x},\omega){h}_{\mu \nu }(\mathbf{x },\omega)\\ 
 & \qquad \qquad \qquad \qquad -\frac{1}{2}\left | {h}_{\lambda }\,^{\lambda }\left ( \mathbf{x},\omega \right )\right |{}^{2}].
\end{aligned}
\end{equation}
In the equation, we've taken the Fourier transform of ${h}_{\mu \nu }(\mathbf{x},t)$ for given $\mathbf{x}$:
\begin{equation}
{h}_{\mu \nu }(\mathbf{x},t )=\int_{-\infty }^{\infty }{h}_{\mu \nu }(\mathbf{x},\omega ){e}^{i\omega t}d\omega.
\end{equation}
We can compute the derivative of ${h}_{\mu \nu }(\mathbf{x},t)$ as follows:
\begin{equation}
\begin{aligned}{h}_{\mu \nu ,0}(\mathbf{x},t )
 & =\int_{-\infty }^{\infty }{{h}}_{\mu \nu ,0}(\mathbf{x},\omega ){e}^{i\omega t}d\omega  \\ 
 & =\int_{-\infty }^{\infty }(-i\omega) {{h}}_{\mu \nu }(\mathbf{x},\omega ){e}^{i\omega t}d\omega.
\end{aligned}
\end{equation}
So, Eq. (\ref{dP}) can be rewritten as
\begin{equation}
\begin{aligned}\frac{dP}{d\Omega }(\mathbf{x },\omega)
 &=\frac{{r}^{2}}{32\pi G} [ {{h}_{,0}}^{\mu \nu \dagger }(\mathbf{x},\omega){{h}}_{\mu \nu ,0}(\mathbf{x },\omega) \\ 
 & \qquad \qquad \qquad \qquad  -\frac{1}{2}\left | {h}_{\lambda ,0}\,^{\lambda }\left ( \mathbf{x},\omega \right )\right |{}^{2} ].
\end{aligned}
\end{equation}
Based on the form of the average power, we can calculate the total energy emitted per unit of solid angle emitted in direction $\mathbf{\hat{x}}$ (stands for the normal vector along $\mathbf{x}$) as follows: 
\begin{equation}
\frac{dE}{d\varOmega }(\mathbf{x})=2\pi  \int_{-\infty }^{\infty}\frac{dP}{d\Omega }(\mathbf{x },\omega)d\omega .
\end{equation}
Plug in the expression for $\frac{dP}{d\Omega }(\mathbf{x },\omega)$,
\begin{equation}
\begin{aligned}\frac{dE}{d\varOmega }(\mathbf{x})
 &= \frac{{r}^{2}}{16G}\int_{-\infty }^{\infty }d\omega [{{h}_{,0}}^{\mu \nu \dagger }(\mathbf{x},\omega){{h}}_{\mu \nu ,0}(\mathbf{x },\omega)\\ 
 & \qquad \qquad \qquad \qquad \qquad -\frac{1}{2}\left | {h}_{\lambda,0 }\,^{\lambda }\left ( \mathbf{x},\omega \right )\right |{}^{2}  ]\\ 
 &=\frac{{r}^{2}}{32\pi G }\int_{-\infty }^{\infty }dt [{{h}_{,0}}^{\mu \nu \dagger }(\mathbf{x},t){{h}}_{\mu \nu,0 }(\mathbf{x },t) \\ 
 & \qquad \qquad \qquad \qquad \qquad -\frac{1}{2}\left | {h}_{\lambda,0 }\,^{\lambda }\left ( \mathbf{x},t\right )\right |{}^{2}  ],
\end{aligned}
\end{equation}
where the second equal sign has used the Parseval's theorem:
\begin{equation}
\frac{1}{2\pi }\int_{-\infty }^{\infty}\left | {h}_{\mu \nu ,0}(t)\right |{}^{2}dt=\int_{-\infty }^{\infty}\left | {h}_{\mu \nu ,0}(\omega )\right |{}^{2}d\omega .
\end{equation}
Then we plug in the specific form of ${h}_{\mu \nu,0}$ for disturbed test masses. When we calculate the total energy emitted, we will just consider the contribution of the leading component ${h}_{03,0}$ and ${h}_{30,0}$:
\begin{equation}
\frac{dE}{d\varOmega }(\mathbf{x})=\frac{{r}^{2}}{32\pi G }\int_{-\infty }^{\infty }\frac{32{G}^{2}}{{r}^{2}}{M}^{2}{a}^{*2}dt,
\end{equation}
Then, the total power of the energy dissipated by the disturbed test mass through its self-gravitational effect is
\begin{equation}
\label{gonglv}
{P}_{GD}=\frac{dE}{dt}=4 G{M}^{2}{a}^{2}.
\end{equation}
This result is similar to the dissipation effect of charged particles (the Larmor formula) in a random electromagnetic field.

Such dissipation effect exists in the other two directions.
If the test mass has accelerations along the other two directions, they will also contribute to dissipation. Then, the total power of energy dissipation is proportional to $\left | \vec{a}\right |{}^{2}$,
\begin{equation}
{P}_{GD}=\frac{dE}{dt}=4 G{M}^{2}\left | \vec{a}\right |{}^{2}.
\end{equation}

\subsection{Generalized Langevin equation for test mass in SGWB}

In the classical diffusion process, the Langevin equation is only suitable for situations where fluctuating forces are white noise, and when we study random dynamical variables driven by colored noise, we need to introduce the generalized Langevin equation \cite{Kubo1991}
\begin{equation}
\label{gl}
M\frac{d{v}_{i}(t)}{dt}=-M\int_{-\infty }^{t}\gamma (t-{t}^{\prime}){v}_{i}({t}^{\prime})d{t}^{\prime}+F{}_{i}(t), \: i=1,2,3.
\end{equation}
For 3D diffusion motion, the random movement of Brownian particles in each direction can be represented by this equation.
In this equation, $\gamma (t)$ represents the friction function and is used to describe the establishment of delayed dissipation. When the autocorrelation function of the fluctuating force is a $\delta $ function, the friction function degenerates to the dissipation coefficient. In addition, The relationship between the spectral density function of velocity in a particular direction and the spectral density function of the fluctuating force in this direction is as follows \cite{Noelle2010} :
\begin{equation}
\label{Sv}
{S}_{{v}_{i}}\left ( \omega \right )=\frac{1}{{M}^{2}}\frac{1}{\left | \gamma \left ( \omega \right )-i\omega \right |{}^{2}}{S}_{{F}_{i}}\left ( \omega \right ),
\end{equation}
where $ \gamma \left ( \omega \right )$ is the Fourier–Laplace transform of $ \gamma \left ( t\right )$.

Inspired by the case of stochastic electromagnetic fields, we should know that not all spectral density functions of $F\left ( t \right )$ can achieve equilibrium with Brownian particles. In the case of stochastic electromagnetic fields, a kind of such spectral density function is known as the black-body radiation spectrum function
\begin{equation}
u\left ( \omega ,T\right )=\frac{\left ( \hbar{\omega }^{3}\right /{\pi }^{2}{c}^{3})}{\left [ exp\left (\hbar\omega /kT \right )-1\right ]},
\end{equation}
where $u\left ( \omega ,T\right )$ is the energy density of the electromagnetic field. 
Therefore, the natural question is that, after long-term evolutions, if the system of Brownian test masses and the SGWB approach equilibrium, what will be the statistical properties of the test masses and the form of the spectral density function of the SGWB?  

In the previous section, we learned that in an SGWB, test masses dissipate energy in a way much like that of charged particles in a stochastic electromagnetic field, so the spectral density function we are looking for should have some similarities with the black-body spectrum. We can get the answer from the law of energy conservation. Let's multiply both sides of the Eq. (\ref{gl}) by ${v}_{i}\left ( t\right )$,
\begin{equation}
\label{conserve}
\frac{M}{2}\frac{d{v}_{i}^{2}\left ( t\right )}{dt}=-M\int_{-\infty }^{\infty}\tilde{\gamma }\left ( t-{t}^{\prime}\right ){v}_{i}\left ( {t}^{\prime}\right ){v}_{i}\left ( t\right )d{t}^{\prime}+F_{i}\left ( t\right ){v}_{i}\left ( t\right ),
\end{equation}
where we have introduced the function $\tilde{\gamma }\left ( t\right )=\Theta \left ( t\right )\gamma \left ( t\right )$, and $\Theta \left ( t\right )$ is the Heaviside function which vanish for $t< 0$. 
Again, like in previous subsections, we will first work out the stochastic motions along one direction ($x^3$-direction), and for simplicity, we will omit the specific subscripts in the following descriptions. 
Now, we assume that test masses reach equilibrium with the SGWB after a long period of random motion. And then, let us average both sides of the equation. $\left \langle F\left ( t\right )v\left ( t\right )\right \rangle$  represents the average work done on the Brownian test mass.
Given $t\gg 0$, one has $M\left \langle v{}^{2}\right \rangle/2={E}_{k}=constant$, so the left side of Eq. (\ref{conserve}) equals 0. The right side represents that the average dissipation power equals the average power of work done by the background.
Then, we get the power of the energy dissipation $\left \langle {P}_{GD}\right \rangle$,
 \begin{equation}
\begin{aligned}\left \langle {P}_{GD}\right \rangle
 &=M\int_{-\infty }^{\infty}\tilde{\gamma }\left ( t-{t}^{\prime}\right ) \left \langle v\left ( {t}^{\prime}\right )v\left ( t\right )\right \rangle d{t}^{\prime} \\ 
 & =M\int_{-\infty }^{\infty}\tilde{\gamma }\left ( \tau \right ){\kappa }_{v} \left ( \tau \right )d\tau ,
\end{aligned}
\end{equation}
where ${\kappa }_{v} \left ( \tau \right )$  is the autocorrelation function of velocity. ${\kappa }_{v} \left ( \tau \right )$ is obviously an even function, so we can make an even extension of $\tilde{\gamma }\left ( \tau \right )$ and $\left \langle {P}_{GD}\right \rangle$ becomes
\begin{equation}
\left \langle {P}_{GD}\right \rangle=\frac{M}{2}\int_{-\infty }^{\infty}{\gamma }\left ( \tau \right ){\kappa }_{v} \left ( \tau \right )d\tau .
\end{equation}
Then, one takes the Fourier transform of ${\gamma }\left ( \tau \right )$ and ${\kappa }_{v} \left ( \tau \right )$ together,
\begin{equation}
\begin{aligned}\left \langle {P}_{GD}\right \rangle
 & =\frac{M}{2}\int_{-\infty }^{\infty}d\tau d{\omega } {d\omega' }\frac{{\gamma }_{f}\left ( {\omega }\right ){S}_{v} \left ( {\omega' }\right )}{(2\pi)^{2}}{e}^{i\left ( {\omega }+{\omega' }\right )\tau }\\ 
 & =\frac{M}{2}\left ( \frac{1}{2\pi }\right )\int_{-\infty }^{\infty}d\omega {\gamma }_{f}\left ( \omega \right ){S}_{v} \left ( \omega \right ),
\end{aligned}
\end{equation}
where ${\gamma }_{f}\left ( \omega \right )$ is the Fourier transforming of ${\gamma }\left ( \tau \right )$ and the Wiener-Khintchine theorem is used to obtain the final result.
The relation between ${\gamma }_{f}\left ( \omega \right )$ and $ \gamma \left ( \omega\right )$ reads
\begin{equation}
{\gamma }_{f}\left ( \omega \right )=2\mathfrak{R}e \gamma \left ( \omega \right ).
\end{equation}
At the same time, one notices that $\left \langle {P}_{GD}\right \rangle$ can also be derived directly through GW radiation back reactions in the previous subsection and  be written as the function of $\left \langle {a}^{2}\right \rangle$ as in Eq. (\ref{gonglv}),  and $a$, being the mean square derivative of $v$, has the spectral density function according to Eq. (\ref{kapa}) and Eq. (\ref{Sv}) 
\begin{equation}
{S}_{a}\left ( \omega \right )=\frac{1}{{M}^{2}}\frac{{\omega }^{2}}{\left | \gamma \left ( \omega \right )-i\omega \right |{}^{2}}{S}_{F}\left ( \omega \right ).
\end{equation}
Therefore, one has
\begin{equation}
\label{PGD}
\begin{aligned}\left \langle {P}_{GD}\right \rangle
 &=4 G{M}^{2}\left \langle {a}^{2}\right \rangle \\ 
 & =4 G{M}^{2}\frac{1}{2\pi }\int_{-\infty }^{\infty }\frac{1}{{M}^{2}}\frac{{\omega }^{2}}{\left | \gamma \left ( \omega \right )-i\omega \right |{}^{2}}{S}_{F}\left ( \omega \right )d\omega .
\end{aligned}
\end{equation}
Comparing the two expressions of $\left \langle {P}_{GD}\right \rangle$, one has
\begin{equation}
4 G{M}^{2}\frac{1}{{M}^{2}}\frac{{\omega }^{2}}{\left | \gamma \left ( \omega \right )-i\omega \right |{}^{2}}{S}_{F}\left ( \omega \right )=M\mathfrak{R}e \gamma \left ( \omega \right ){S}_{v}\left ( \omega \right ),
\end{equation}
\begin{equation}
4 G{M}^{2}{\omega }^{2}{S}_{v}\left ( \omega \right )=M\mathfrak{R}e \gamma \left ( \omega \right ){S}_{v}\left ( \omega \right ),
\end{equation}
which gives rise to the following important relation for SGWB at equilibrium with the test mass system 
\begin{equation}
\label{haosan}
\mathfrak{R}e \gamma \left ( \omega \right )=4 GM{\omega }^{2}.
\end{equation}
This conclusion results from the second law of thermodynamics and the equipartition theorem of energy. 
Here, we introduce a cut-off factor ${f}_{k}^{2}$, which satisfies the condition that equaling 1 up to some large frequency $\mathcal{W}$ and falling to zero thereafter. A convenient form that satisfies this condition is \cite{Ford1988}
\begin{equation}
{f}_{k}^{2}=\frac{{\mathcal{W} }^{2}}{\omega {}^{2}+{\mathcal{W} }^{2}}.
\end{equation}
Then, we can multiply $\mathfrak{R}e \gamma \left ( \omega \right )$ by ${f}_{k}^{2}$
\begin{equation}
\label{wentai}
\mathfrak{R}e \gamma \left ( \omega \right )=4 GM{\omega }^{2}{f}_{k}^{2}.
\end{equation}
We notice that if Eq. (\ref{haosan}) is applicable across all frequencies, this implies $\mathcal{W} \to \infty $ and ${f}_{k}^{2}=1$ across all frequencies.
In this case, we have:
\begin{equation}
\gamma \left ( t\right )=4GM\ddot{\delta }\left ( t\right ).
\end{equation}
Then, the Fourier–Laplace transform of $ \gamma \left ( t\right )$ is
\begin{equation}
\begin{aligned}\gamma \left ( \omega \right )
 & =\int_{0}^{\infty }4GM\ddot{\delta }\left ( t\right ){e}^{i\omega t}dt\\ 
 & =2GM{\omega }^{2}
\end{aligned}
\end{equation}
According to the generalized Langevin model, when in equilibrium, the form of the fluctuation-dissipation theorem is
\begin{equation}
\label{zhanghao}
\mathfrak{R}e \gamma\left ( \omega  \right )=\frac{1}{2MkT}{S}_{F}\left ( \omega  \right ),
\end{equation}
where $\mathfrak{R}e \gamma\left ( \omega \right )$ and ${S}_{F}\left ( \omega  \right )$ differ only by a constant factor $2MkT$. Substitute ${S}_{F}\left ( \omega  \right )$ and $\gamma \left ( \omega \right )$ into the Eq. (\ref{PGD}). One will find that $\left \langle {P}_{GD}\right \rangle$ becomes infinity in this case. This is not by physical reality. So, $\mathcal{W} \to \infty $ represents a condition that does not conform to physical facts, and the introduction of a cut-off factor ${f}_{k}^{2}$ is necessary. This implies that the classical version of the equipartition theorem may not hold at high frequencies. 

Based on Eq. (\ref{haosan}), for a finite $\mathcal{W}$,
\begin{equation}
\label{haosan2}
\gamma \left ( t\right )=4GM{\mathcal{W} }^{2}\left [ \delta \left ( t\right )-\mathcal{W} e^{-\mathcal{W} \left | t\right | }\right ].
\end{equation}
The same derivation can be carried out in all three directions and $\gamma \left ( t\right )$ can be adaptable to all the components of Eq. (\ref{gl}). In the 3D case, the fluctuation-dissipation theorem becomes
\begin{equation}
\mathfrak{R}e \gamma\left ( \omega  \right )=\frac{1}{6MkT}{S}_{\left | \vec{F}\right |}\left ( \omega  \right ).
\end{equation}

Historically, there have been similar situations in the study of the BBR field. 
Here, Eq. (\ref{haosan}) is the analog to the famous Rayleigh-Jeans spectrum. Rayleigh and Jeans derived this spectral function through the theorem of energy equipartition. However, if it were to hold at all frequencies, for classical theories, the well-known ultraviolet catastrophe would take place. The breakdown of the principle of energy equipartition at high frequencies indicates that we cannot regard GW as a classical wave across all frequencies. The establishment of a consistent quantum theory for GW is imperative.

We should know that the physically significant results of a long evolution for this model should not depend on the details (except $\mathcal{W} $) of the cut-off factor because the results of long-time evolutions should be determined by the characteristics of $\mathfrak{R}e \gamma \left ( \omega \right )$ at low frequencies.
In addition, according to Eq. (\ref{zhanghao}), we can conclude that only when ${S}_{\left | \vec{F}\right |}\left ( \omega  \right )$ satisfies the condition of being proportional to the square of ${\omega}^{2}$ at low frequencies, which can be called the equilibrium condition, can a test mass have a final average kinetic energy that does not change with time. This conclusion is closely related to Eq. (\ref{gonglv}). 
Furthermore,  the effective temperature $T$ of the SGWB naturally emerged from the above analysis. 
Given the SGWB that satisfies the equilibrium condition, based on Eq. (\ref{sf}), we can obtain the spectral density function of ${h}_{\mu \nu }$,
\begin{equation}
{S}_{ij}\left ( \omega \right )=S{}\left ( \omega \right )={S}{f}_{k}^{2},
\end{equation}
where $S$ is a constant.
Then, ${S}_{\left | \vec{F}\right | }$ can be written as 
\begin{equation}
{S}_{\left | \vec{F}\right |}\left ( \omega \right )=\frac{11}{4}{M}^{2}S{\omega }^{2}{f}_{k}^{2}.
\end{equation}
According to Eq. (\ref{zhanghao}), the effective temperature of this macroscopic system can be obtained as
\begin{equation}
\label{temp}
T=\frac{1}{32G{k}_{B}}\frac{11S}{12}.
\end{equation}
The effective temperature $T$ is only related to the spectral intensity of the SGWB and is independent of the mass parameters of the test objects. This undoubtedly conforms to the second law of thermodynamics. All test masses tend to approach the same final kinetic energy and satisfy the equipartition theorem in the low-frequency region. In the study of classical Brownian motion, this is a natural result, since the temperature is a property of the thermal background. However, for SGWB, this turns out to be an enlightening result, since it establishes in some sense the deeper compatibility of GR with the laws of thermodynamics.
Considering the similarity between the energy spectrum function of the SGWB and ${S}_{\left | \vec{F}\right |}\left ( \omega  \right )$, the limitation of equilibrium conditions on ${S}_{\left | \vec{F}\right |}\left ( \omega  \right )$ can be described as the limitation on the energy spectrum function of the SGWB. 

In addition, we can conclude that only if the energy spectrum function of the SGWB satisfies the equilibrium conditions Eq. (\ref{wentai}), its entropy should reach the maximum (for a reasonable definition of the entropy of the SGWB that satisfies the second law of thermodynamics). 
If the energy spectrum function of the SGWB does not satisfy the equilibrium conditions, the test masses in it will not reach a final constant kinetic energy. This implies that the kinetic energy of the test mass will never reach a fixed value. Then, test masses will permanently exchange energy with the background, and the entropy production rate can never be zero.

As for the details of the diffusion process, based on the generalized Langevin equation, we get the time-dependent diffusion coefficient  $D (t)$,
\begin{equation}
D\left ( t\rightarrow \infty \right )=\lim_{t\to\infty } \frac{1}{6}\frac{d}{dt}\left \langle \Delta {\left | \vec{x}\right |}^{2}\left ( t\right )\right \rangle=\frac{{k}_{B}T}{M}b\left ( 1-b\right )t,
\end{equation}
where $b$ is the non-Markovian influence factor
\begin{equation}
b=\frac{1}{1+4GM\mathcal{W} }.
\end{equation}
The mean and variance of velocity are expressed as follows:
\begin{gather}
\left \langle \vec{v}\left ( t\rightarrow \infty \right )\right \rangle=b\vec{v}\left ( 0\right ),
\\\left \langle \left | \vec{v}\right |{}^{2}\left ( t\rightarrow \infty \right )\right \rangle=\frac{3{k}_{B}T}{M}+{b}^{2}\left [ {\left | \vec{v}\left ( 0\right )\right |}^{2}-\frac{3{k}_{B}T}{M}\right ].
\end{gather}
The factor $b$ determines the non-Markovian nature of this process and the final average kinetic energy for testing quality will also be affected by its own mass because of the non-Markovian nature. For a fixed $\mathcal{W}$,
if $4GM\mathcal{W}\ll 1$, which means that $M$ is not large enough, the test mass will be dominated by its initial conditions and not display Markovic nature. On the contrary, if $4GM\mathcal{W}\gg 1$, the test mass will move as the so-called “Markovic motion”.
One can see that the larger the mass of a test mass, the closer its non-Markovian factor $b$ to zero, which means it tends to forget its initial information and reach a temperature that is exactly the same as the background field.

According to the above relation between the diffusion coefficient and time, we can conclude that the variance of the diffusion distance is proportional to the square of time, $\left \langle {\left | \vec{x}\right |}^{2}\right \rangle\propto {t}^{2}$, meaning that the anomalous diffusion exponent $\alpha =2$. This is a type of superdiffusion, or it can also be referred to as ballistic diffusion.

\section{Conclusions}

We investigated diffusive motion in an SGWB. Based on the development of the Langevin equation in two classical models of diffusive motion, we performed a stochastic dynamical analysis of a test mass undergoing random motion in the SGWB. From this analysis, we derived a generalized Langevin equation that governs this system.

In this paper, a core result is the provision of the energy dissipation behavior of test mass in the cosmological rest frame of the SGWB, Eq. (\ref{gonglv}). We are surprised by the striking similarity of this result to the energy dissipation behavior of charged particles in a stochastic electromagnetic field. When we plug this result into the theory of stochastic dynamics, such a form of dissipated energy, $P\propto{a }^{2}$, will lead to a power spectrum of the background fluctuating forces at equilibrium with the form, ${S}_{\left | \vec{F}\right |}\left ( \omega \right )\propto{\omega  }^{2}$, at low frequencies. The latter, for a classical stochastic wave field, is essentially an expression of the principle of energy equipartition. Each wave mode possessing the same average energy is known as the Rayleigh-Jeans spectrum. The results from the two theories, however, are well unified within the framework of the generalized Langevin equation. Based on this result, it is reasonable to have such a dissipative form for test masses when we consider gravitational waves as a kind of classical wave. 

Using the fluctuation-dissipation theorem, we then obtain the temperature $T$ of an SGWB, Eq. (\ref{temp}). This temperature is merely a property of the background and is the same for different test masses moving within the background. For a stochastic background, temperature is a very important macroscopic quantity. Based on the definition of temperature, we can further explore more thermodynamic quantities, which will also be our future work.

Pulsar timing array (PTA) data is already a well-established technique for studying SGWB. Our next step is to use PTA to investigate the observable effects of our model, hoping to find evidence supporting our model from existing data.

Additionally, just as the blackbody radiation of the electromagnetic field inspired the establishment of quantum mechanics, we believe that the thermodynamic study of the SGWB will definitely be of great significance for further understanding the nature of gravity. For instance, as in the results presented in our article, macroscopic steady states only impose restrictions on the form of the background power spectrum at low frequencies. What kind of power spectrum shape should the high-frequency part of a steady SGWB have, and whether it will satisfy the Bose-Einstein distribution? This question cannot be answered through macroscopic processes. However, at least in the low-frequency case, our results do not violate the Bose-Einstein distribution rate.

Another interesting direction is the relationship between the IO model and the SGWB. In the case of a BBR field, we can fully describe the random motion of a charged particle through the IO model. Given the similarity between the SGWB and the BBR scenario, it might be possible to find a model analogous to the IO model under these circumstances. We believe that an in-depth study of the thermodynamics of gravitational waves could offer new perspectives on gravity.

Finally, the impact of ballistic diffusion behavior on the distribution of matter in the universe is also an interesting area of research.

\section{acknowledgment}

This work is supported by the National Key Research, Development Program of China, Grant No. 2021YFC2201901, and the International Partnership Program of the Chinese Academy of Sciences, Grant No. 025GJHZ2023106GC.

\bibliography{references}

\end{document}